\documentclass{elsart}
\usepackage{epsfig}
\begin{document}
\begin{frontmatter}
\title{Wolf-Rayet Stars in Starburst Galaxies\thanksref{CICYT}}
\author{J. Miguel Mas-Hesse}
\address{LAEFF-INTA, Apdo. 50727, 28080 Madrid, mm@laeff.esa.es}
\author{Daniel Kunth}
\address{IAP, 98bis Bd. Arago, 75014 Paris, kunth@iap.fr}
\author{Miguel Cervi\~no\thanksref{ESA}}
\address{Observatoire Midi-Pyr\'en\'ees, 14, Av. E. Belin, 31400 Toulouse,
  mcervino@ast.obs-mip.fr} 

\thanks[CICYT]{Work supported by Spanish CICYT under grant ESP95-0389-C02-02}
\thanks[ESA]{ESA fellow}

\begin{abstract}

Wolf-Rayet stars have been detected in a large number of galaxies
experiencing intense bursts of star formation. All stars initially more
massive than a certain, metallicity-dependent, value are believed to
experience the Wolf-Rayet phase at the end of their evolution, just before
collapsing in supernova explosion. The detection of Wolf-Rayet stars puts
therefore important constraints on the evolutionary status of starbursts,
the properties of their Initial Mass Functions and their star formation
regime. In this contribution we
review the properties of galaxies hosting Wolf-Rayet stars, with special
emphasis on the factors that determine their presence and evolution, as
well as their impact on the surrounding medium. 

\end{abstract}
\end{frontmatter}

\section{Introduction}

In the last 20 years, Wolf-Rayet stars have been detected in several
extragalactic objects. Allen et al. (1976) \cite{Aletal76} identified for
the first time the characteristic He~II $\lambda$4686 broad atmospheric
emission line in He~2-10. Conti (1991) \cite{Con91} listed already 37
objects showing WR features, a number which was increased to more than 130
in Schaerer and Vacca (1999) \cite{SV98}, and which is continuously
increasing. The WR features are broad, but generally weak, so that they can
be detected only in spectra with high signal to noise in the
continuum. This explains why they were not identified in the first years of
emission line galaxies spectroscopy. The detectors narrow dynamical range
prevented having good signal to noise simultaneously on the bright emission
lines and in the weak continuum of these galaxies.  More careful searches
in the last years have allowed to also identify the broad feature around
CIV $\lambda$5808 attributed to WC stars, a subtype of WR stars
characterized by strong and broad C emission lines. We show in Fig.~1 the
optical spectrum of IZw~18, with the identification of some typical WR
lines.

\begin{figure}  
\begin{center}\psfig{figure=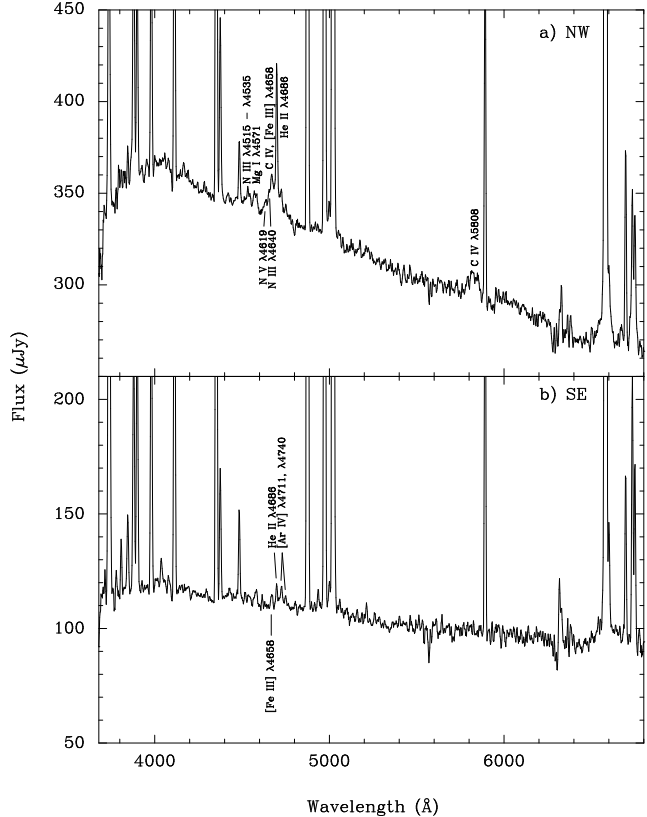,height=11cm}\end{center}
\caption{Optical spectrum of IZw~18 NW and SE knots,  showing the typical
  WR features around HeII $\lambda$4686 and around CIV $\lambda$5808
  (adapted from Izotov et al. (1997) \cite{Izetal97}).}
\end{figure}

Wolf-Rayet stars have been found in very different extragalactic
environments: Giant HII regions, Blue Compact galaxies, generic emission
line galaxies, IRAS galaxies, Seyfert galaxies,...., in general, always in
regions experiencing a strong episode of massive star formation. This fact
provided in the 80's a definitive support to the so-called ``Conti
scenario'', according to which WR stars were the descendants of massive
stars, experiencing this short evolutionary phase (around 500.000 years)
just before collapsing into a supernova explosion. Conti (1991)
\cite{Con91} successfully proposed the term ``Wolf-Rayet galaxies'' to
group all the galaxies hosting WR stars. Given the very different kind of
objects in which WRs have been identified and the fact that their detection
depends in most cases just on the observational strategy, this term has to
be taken with care. IZw~18 could be a prototype for that. Being the most
metal deficient galaxy known, it has been observed for years aiming to
measure in detail the intensities of the different emission lines, with no
WR feature being detected at all. However, long integrations on big
telescopes allowed two independent groups to identify in 1997 the WR
features around HeII and CIV, adding this object to the WR galaxy list
(Izotov et al. 1997 \cite{Izetal97}, Legrand et al. 1997 \cite{Leetal97}).

In this contribution we will review the parameters that control the
presence of Wolf-Rayet stars in star-forming environments, as well as their
effects on the surrounding medium.  Our goal will be to summarize what the
detection of Wolf-Rayet stars can tell us about the properties of 
massive star-formation episodes in different environments.

\section{What can we learn from the presence of WR's?}

The detection and quantification of the number (and type!) of Wolf-Rayet
stars, and their ratio to OB stars, provide a bulk of information about the
intrinsic properties of the different star formation episodes, like Initial
Mass Function slope and limits, star formation regime, and so on. Let's
summarize first which factors drive the formation of WR's in star-forming
environments. 

\subsection{What controls the formation of WR'S?}

As predicted by present stellar evolutionary tracks, there are mainly three
parameters controlling the formation of WR's: 

\begin{itemize}

\item Metallicity. 

\item Initial Mass Function (IMF) limits. 

\item Properties of binary systems. 

\end{itemize}

The Wolf-Rayet phase is characterized by the ejection via strong stellar
winds of the outer layers of evolved massive stars. The efficiency in
powering these winds is clearly a function of the metallicity, so that the
lower the metallicity, the higher the initial mass required for a star to
become a WR. The precise values for this mass limit depends also on the
mass loss rate prescriptions and the rotation properties of a given star.
Following a conservative mass loss rate scenario, Mas-Hesse and Kunth 1991
\cite{MHK91} and Cervi\~no and Mas-Hesse 1994 \cite{CMH94} estimated the
lower mass limit for WR formation at solar metallicity to be 32 M$_\odot$.
Lower and more realistic mass limits are reached if the mass loss rate are
somewhat enhanced, as discussed in (Schaerer and Vacca 1998
\cite{SV98}). In general we can say that a star will become a WR if its
initial mass is above 20 M$_\odot$ for solar metallicity, and above 80
M$_\odot$ at Z~=~Z$_\odot$/10.  Therefore, the detection of a significant
number of WR stars in low metallicity environments, as in IZw~18, directly
implies that the upper mass limit of the IMF has to be close to 100
M$_\odot$.

The evolution of massive stars in binary systems can also lead to the
formation of WR's.  Around 50\% of massive stars are believed to form in
binary systems, out of which around 5\% are expected to evolve as massive
close binaries. Such close binaries experience different processes of mass
transfer during their evolution, which can lead to the formation of WR
stars at ages where no WRs would exist according to the evolution of single
stars (Cervi\~no 1998 \cite{Cer98}, Vanbeveren 1998 \cite{Van98}and
references therein, Cervi\~no
et al. 1999 \cite{Ceetal99}). First, a star can loose completely its outer
envelope at the end of the H burning phase, with a naked core emerging
which could have very similar properties to single WR stars. Second,
accretion of mass would allow a star of initial medium/low mass to evolve
as an initially massive star, becoming a WR at late evolutionary stages of
the starburst. Summarizing, while the standard Conti scenario predicts the
presence of WR stars only between 2 and 6 Myr after the onset of the burst
(only 3 to 4 Myr at low metallicities!), the binary channel predicts a
rather constant amount of WR stars between 5 and around 20-30 Myr, as shown
in Figs.~2 and 3.

\begin{figure} 
\begin{center}\psfig{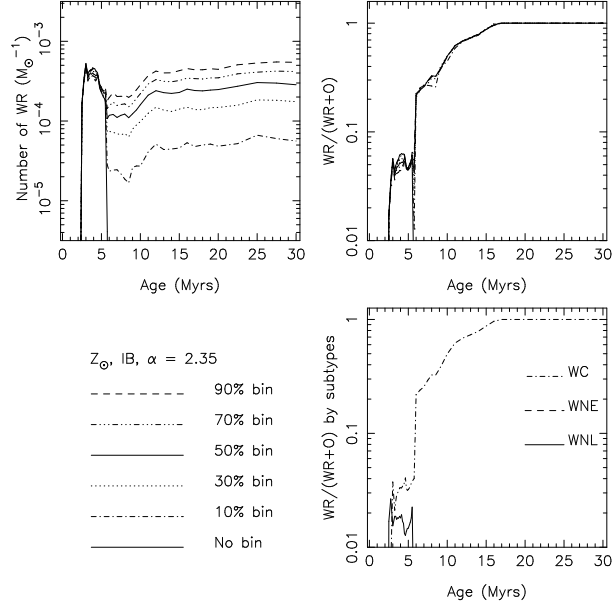}\end{center}
\caption{Predicted number of WR stars as a function of the abundance of
  binary systems. All WRs appearing after 6~Myr are formed by mass transfer
  processes in massive close binaries. More details are given in Cervi\~no
  et al. (1999) \cite{Ceetal99}.}
\end{figure}

\begin{figure}  
\begin{center}\psfig{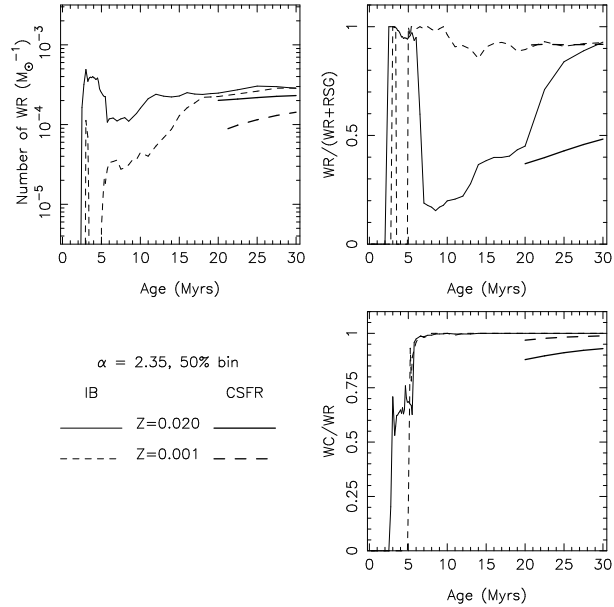}\end{center}
\caption{Predicted population of Wolf-Rayet stars as a function of
  metallicity, assuming 50\% of binary systems and Salpeter's IMF. }
\end{figure}

\subsection{WR features detectability}

In the previous section we have summarized the parameters affecting the
presence of WR stars at a given time during a star formation episode. But
even if they are present, their detection and quantification is furthermore
affected by some additional questions: 

\begin{itemize}

\item Star formation regime. 

\item Underlying stellar population. 

\item Differential reddening. 

\end{itemize}

It has been well established that the present star formation rates showed
by several starbursting galaxies can not have been maintained over long
periods of time without exhausting the estimated original amounts of
gas. It seems that massive star formation proceeds in these objects as
(maybe repeated) short-lived, very intense episodes. The question now is
how short are really these episodes: almost instantaneous or extended over
tens of million years? Several arguments point towards almost coeval star
formation, i.e., all stars (at least, all massive stars) would have been
formed almost simultaneously, or in any case within few million years. The
ignition of hundreds or thousands of massive stars within relatively small
volumes and within relatively short times would probably inhibit the
further formation of stars, at least for several million years, until the
most massive stars start to fade out.

The detection of Wolf-Rayet stars provides important constraints on this
issue. In extended star formation scenarios massive stars would be
continuously formed during tens of million stars. Since the WR phase lasts
for only around 500.000 years, the net effect is that the expected
$L(WR)/L(H\beta)$ ratio would be significantly smaller than for coeval
starbursts. We show in Fig.~4 the predictions for this ratio at different
metallicities and assuming different IMF slopes, both for an instantaneous
burst and for an extended star formation episode. We have plotted on the
figures the mean vaues compiled by Mas-Hesse and Kunth (1999)
\cite{MHK99}. It can be seen that, first, the distribution of observed
values fall rather well within the predictions of coeval models, while,
second, the observations are barely consistent, at most, with the
predictions of extended star formation episodes. We can conclude,
therefore, that the formation of massive stars in Wolf-Rayet galaxies
proceeds almost coevally, in any case within few million years.

\begin{figure} 
\begin{center}\psfig{figure=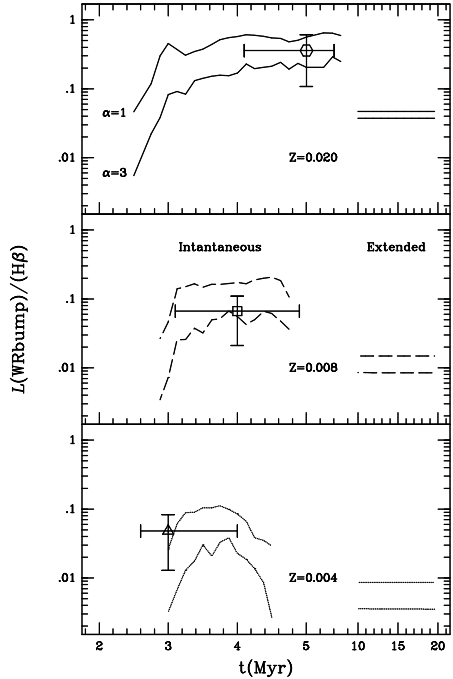,height=9cm}\end{center}
\caption{Predicted $L(WR)/L(H\beta)$ ratio for different metallicities and
  two extreme IMF slopes. The data points correspond to the compilation of
  Mas-Hesse and Kunth (1999) \cite{MHK99}.  }
\end{figure}

Another factor strongly affecting the detectability of the WR features on
the stellar continuum spectra is the presence of an underlying, older
stellar population. Up to now, WR stars have been generally detected in
galaxies whose optical continuum is mostly dominated by the newly formed,
massive stars, with older stars contributing less than 50\% to the total
continuum at around 5000~\AA\ (see the examples in Mas-Hesse and Kunth
1999 \cite{MHK99}). 
But if a starburst takes place in a galaxy with an
important older stellar population, the WR features would be diluted within
the optical continuum, and would be harder to be detected.

Finally, two additional factors can lead to significant errors in the
quantification of the relative WR vs. OB stars population as derived from
the observed $L(WR)/L(H\beta)$ ratio. First, it has to be taken into
account that the $L(H\beta)$ emission is spread over a relatively large
area ionized by the cluster of young, massive stars. On the other hand, the
WR features are associated to the stellar population, and are therefore
spatially restricted within a much smaller region. Therefore, if
$L(WR)/L(H\beta)$ is derived from single narrow slit observations, the
ratio can be severely overestimated, since it would have been contributed
by most WR stars in the region, but only a fraction of the total H$\beta$
flux. Mas-Hesse and Kunth (1999) \cite{MHK99} have estimated that this
problem can distort the derived ratios by even an order of magnitude,
making them useless for comparison with theoretical predictions. And
second, it has been established in the last years that the extinction
affecting the stellar continuum (and therefore the WR features) might be in
some cases significantly smaller than the extinction affecting the Balmer
emission lines (Schaerer and Vacca 1998 \cite{SV98}). Ma\'\i z-Apell\'aniz
et al. (1999) \cite{Maetal99} showed the spatial decoupling of stars, gas
and dust in the star-forming regions of NGC~4214, which are rich in WR
stars. It seems that the stellar winds can be very efficient in
some cases in blowing away both the nebular gas and the dust grains,
leaving the massive stellar cluster within relatively dust-free volumes. On
the other hand, dust particles were detected mixed with the nebular gas,
yielding relatively large extinctions on the Balmer emission
lines. Ma\'{\i}z-Apell\'aniz et al. (1999) \cite{Maetal99} estimated that
this effect could yield to an overestimation of the observed
$L(WR)/L(H\beta)$ ratio by a factor between 2 and 5.

We conclude therefore that the quantification of the relative number of WR
over OB stars in star-forming regions can be severely overestimated by
different effects. Therefore, reliable constraints on the properties of the
star formation episodes can only be derived when different observational
parameters are analyzed simultaneously, including $L(WR)/L(H\beta)$,
$W(H\beta)$, $EW(WR)$,... as discussed in more detail by Mas-Hesse and
Kunth (1999) \cite{MHK99}.
 
\subsection{Effects of Wolf-Rayet stars on the surrounding medium}

As we have commented above, WR stars appear as the effect of strong stellar
winds blowing out the outer atmospheric layers of evolved massive
stars. The detection of WR's traces therefore the presence of clusters rich
in very massive stars, which are significantly affecting their surrounding
interstellar medium in many ways: 

\begin{itemize}

\item Large amounts of mechanical energy are being injected into the
  medium, even before the production of the first supernova explosions
  after he onset of the burst. Leitherer et al. (1995) \cite{Leetal95} and
  more recently Cervi\~no et al. (1999) \cite{Ceetal99} have evaluated the
  amount of mechanical energy released by these powerful winds. It would be
  enough to blow out the surrounding nebular gas, leading to an empty
  cavity free of gas and dust. Kunth et al. (1998) 
  \cite{Kuetal98} detected outflowing
  gas apparently powered by the central starburst in a number of
  galaxies. Fig.~5 shows the profile of the Ly$\alpha$ emission line,
  clearly absorbed at the blue wing by neutral gas moving at several
  hundreds of km/s. The mechanical energy released would imply that the
  chemical enrichment associated to a new generation of stars wouldn't
  become evident inmediately, since the enriched gas could be thrown away
  to relatively large distances by these gas outflows, as proposed by
  different authors in the last years
  (see the contribution from G. Tenorio-Tagle in this volume).

\begin{figure} 
\begin{center}\psfig{figure=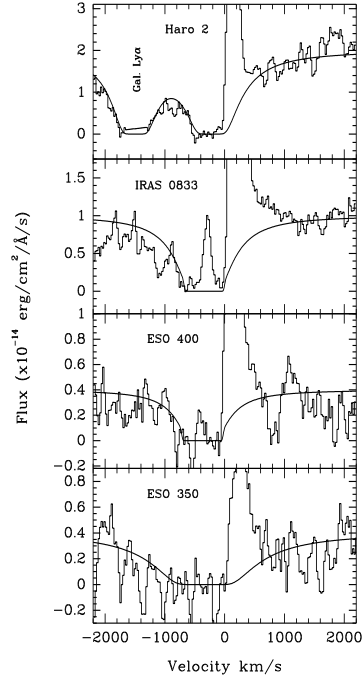,height=9cm}\end{center}
\caption{Lyman alpha profiles of 4 starburst galaxies analyzed by Kunth et
  al. (1998) \cite{Kuetal98}. The effect of neutral gas outflowing at few
  hundreds km/s is evident. }
\end{figure}

\item When a star enters the Wolf-Rayet phase, its naked He core at a very
high effective temperature (around 100.000 K) can become visible, producing
so a source of rather hard ionizing radiation, much harder than the
ionizing flux associated to Main Sequence OB stars (below 50.000 K in any
case). This hard ionizing flux can produce a number of emission lines not
usually found in HII regions. Schaerer (1996) \cite{Sch96} proposed that
some kind of WR stars could provide enough hard ionizing photons to explain
the narrow HeII $\lambda$4686 detected in some, but not in all, starburst
galaxies.

\end{itemize}

\section{Summary and conclusions}

The identification of Wolf-rayet stars in starburst environments in the
last 20 years has helped to place strong constraints on the properties of
these massive star formation episodes. We know presently that these
starbursts are apparently short-lived (all massive stars are essentially
coeval), that the Initial Mass Function in these regions is almost always
close to Salpeter's one (with slope $\alpha = 2.35$), with stars of initial
masses around 100~M$_\odot$ at least. Most of these starbursts formed their
massive stars less than around 6~Myr ago, but this is probably a selection
effect, since after this age the ionizing flux fades rapidly and the
objects do not loook like ``emission line galaxies'' any longer. There are
nevertheless a number of questions still open:

\begin{itemize}

\item The starbursts in which WR stars have been detected seem to have been
  generally very short-lived. But, can we extrapolate this conclusion to
  all starbursts, including those in which no WR stars have been (yet)
  detected?

\item What are the constraints that can be derived from the WN/WC ratios
  observed in different galaxies?  

\item How does rotation affect the predictions of the synthesis models used
  up to now? A. Maeder provides in this volume a summary of the state of
  the art evolutionary tracks including stellar rotation. 

\item Are there really Wolf-Rayet stars at evolved stages of the cluster,
  when the emission line strengths are very small, as predicted by the
  models including the evolution of binary systems? Would these WR's show
  the same features as WR stars formed along the Conti scenario?   

\end{itemize}

Let's continue searching for Wolf-Rayet stars in different environments in
order to help solve these open questions in the near future.

\end{document}